\documentclass[12pt,preprint]{aastex}
\slugcomment{Ver.2 01-Jun-12}

\shorttitle{Far-IR transparent contact}
\shortauthors{suzuki et al.}

\begin{document}


\title{Molecular-beam epitaxial growth of a far-infrared transparent
electrode for extrinsic Germanium photoconductors}


\author{Toyoaki Suzuki, Takehiko Wada, Kazuyuki Hirose, Hironobu Makitsubo\altaffilmark{1}}

\affil{Institute of Space and Astronautical Science, Japan Aerospace
	      Exploration Agency, 3--1--1 Yoshinodai, chuou, Sagamihara,
              Kanagaawa 252--5210, Japan}
\altaffiltext{1}{Department of Astronomy, Graduate School of Science,
	      the University of Tokyo, 7--3--1 Hongo, Bunkyo-ku, Tokyo
	      113--0033, Japan}
\email{suzuki@ir.isas.jaxa.jp}

\and 

\author{Hidehiro Kaneda}
\affil{Graduate School of Science, Nagoya University, Chikusa, Nagoya 464--8602, Japan}

%\and

%\author{Kentaroh Watanabe}
%\affil{Research Center for Advanced Science and Technology, University
%of Tokyo, Meguro, Tokyo 153--8904, Japan}


%\altaffiltext{1}{Visiting Astronomer, Cerro Tololo Inter-American Observatory.
%CTIO is operated by AURA, Inc.\ under contract to the National Science
%\altaffiltext{2}{Society of Fellows, Harvard University.}
%\altaffiltext{3}{present address: Center for Astrophysics,
%    60 Garden Street, Cambridge, MA 02138}
%\altaffiltext{4}{Visiting Programmer, Space Telescope Science Institute}
%\altaffiltext{5}{Patron, Alonso's Bar and Grill}


\begin{abstract}
We have evaluated the optical and electrical properties of a far-infrared (IR)
transparent electrode for extrinsic germanium (Ge) photoconductors at 4
 K, which was fabricated by molecular beam epitaxy (MBE).   
As a far-IR transparent electrode, an aluminum (Al)-doped Ge layer is formed
at well-optimized doping concentration and layer thickness in terms of
 the three requirements: high far-IR transmittance, low resistivity, and
 excellent ohmic contact. The Al-doped Ge layer has the far-IR
 transmittance of $>95$ \% within the wavelength range of 40--200 $\mu$m, while low
 resistivity ($\sim$5 $\Omega$ cm) and ohmic contact are ensured at 4 K.   
We demonstrate the applicability of the MBE technology in
 fabricating the far-IR transparent electrode satisfying the above requirements. 
\end{abstract}



%\keywords{globular clusters: general --- globular clusters: individual(NGC 6397,
%NGC 6624, NGC 7078, Terzan 8}



\section{Introduction}
Bulk germanium extrinsic photoconductors (Ge PCs) have been
widely used in far-infrared (far-IR) astronomical observations in the
50--120 $\mu$m wavelength range~\citep[e.g.][]{haegel1985}. The spectral response
of p-type Ge PCs can be extended to longer wavelengths ($\sim$200
$\mu$m) by applying a uniaxial compressive stress
\citep{kazanskii1977}. In order to provide high quantum efficiency,
broad spectral response, and low cosmic-ray
hitting rates compared to those for the bulk Ge PCs, germanium blocked
impurity band (Ge BIB) detectors have been developed as next-generation
far-IR detectors replacing the bulk Ge PCs \citep{watson1988, wu1991, watson1993, 
huffman1993, bandaru2002, beeman2007, kaneda2011}. 

\begin{figure}
\center{
\hspace{-20mm}
\includegraphics[scale=.80]{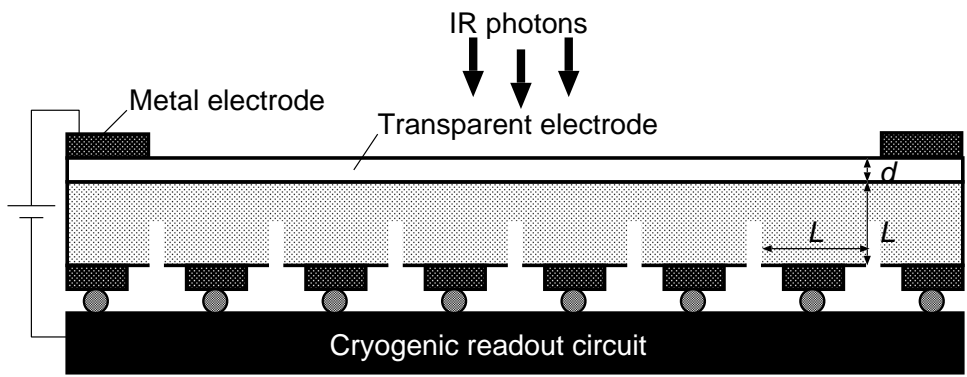}
\caption{Schematic cross-sectional view of a monolithic PC array.}
\label{fig1}
}
\end{figure}

BIB detectors require electrodes with not a transverse but a
longitudinal, and thus transparent configuration to obtain high
responsivity, because a far-IR absorbing thin layer ($\la10$ $\mu$m) is
formed underneath an electrode. And, the construction of a large-format
array for both PCs and BIB detectors is indispensable to improve
observational efficiency and measurement accuracy. The array as shown in
Fig.~\ref{fig1} also needs electrodes with the far-IR transparent
configuration. Although such transparent electrodes have been applied to
Ge PC arrays, performance of transparent electrodes is not
optimized~\citep[eg.][]{fujiwara2003}.    

%In particular, the advantage of
%wavelength response as long as 200 $\mu$m without the use of large
%stressing mechanisms drives constructing a large-format array camera,
%which is important, especially for space astronomical observations, to
%improve observational efficiency and measurement accuracy. Ge PCs that
%have a transparent electrode are suitable for fabricating a large format
%monolithic two-dimensional array as shown in Fig.~\ref{fig1}. 

The transparent electrode requires high far-IR transmittance, while it
must ensure high conductivity and an excellent ohmic contact at temperatures 
below 4 K. %, which are constrained by significantly reducing thermal
%current of Ge PCs. 
To obtain high far-IR transmittance, a thin
layer with a low doping concentration is desirable. On the other hand, 
a low resistivity layer equipped with excellent ohmic contact can be 
achieved by a degenerately doped ($n_\mathrm{d}\sim10^{16}$--$10^{17}$
cm$^{-3}$), thick, and epitaxially grown Ge layer. Thus, the doped Ge
layer should be epitaxially grown with optimally-controlled doping
concentration and thickness.         

Most commonly used technique to fabricate the transparent electrode is
ion implantation on one surface of a device. However, its doping
concentration often shows a graded profile. Therefore,
the formation of the decreasingly-doped layer as a function of implant
depth is inevitable.
%Therefore, the formation of the exceedingly doped layer is inevitable in addition to
%the optimally doped Ge layer. 
Such a broad doping profile significantly degrades far-IR transmittance
due to photon absorption within the thick-implanted layer by
free carriers and doped impurities \citep{hadek1985,
fujiwara2000}. Nevertheless, there are few investigations in terms of
abruptness for the doping profile.  

To improve
far-IR transmittance, an alternative process technology with precise
control of a doping profile, a doping concentration and a layer
thickness should be introduced. A promising candidate technology is
molecular beam epitaxy (MBE) that is an Ultra-High-Vacuum (UHV)-based
technique for producing high quality epitaxial structures with
monomolecular layer (monolayer) thickness control. MBE technology
allows us to accurately control doping profiles, doping concentrations,
and interfaces, thanks to the lower growth rate and temperature compared
to other epitaxial growth technologies such as liquid phase epitaxy. In
this paper, we show the properties of the far-IR transparent
electrode fabricated by the MBE technology.

%  a degenerately doped Ge layer
% is formed as the transparent electrode.  

%  that is a degenerately doped Ge layer. Such layer
%  is essential to 

%Properties required for transparent electrode are
%high transmittance in far-IR region, a good ohmic contact, and high
%electrical conductivity at temperatures below 4.2 K. 

%stressed Ge detectors is the difficulty in fabricating a stress
%mechanism for large to-dimensional arrays.

%especially for space application, 

\section{Requirements for a far-IR transparent contact}
In this section, we search optimized solutions for
the doping concentration $n_\mathrm{d}$ and the layer thickness $d$,
considering the following requirements: high far-IR transmittance, low
resistivity, and excellent ohmic contact. Note that we assume all the
dopants are electrically activated; 
$n_\mathrm{ad} = n_\mathrm{d}$, where $n_\mathrm{ad}$
is activated doping concentration.    

\vspace{-5mm}  
\subsection{Far-IR transmittance}
The far-IR transmittance $T_{\mathrm{IR}}$ through a sample is defined as
$T_{\mathrm{IR}}=I_t/I_0$, where $I_0$ and $I_t$ are intensities of 
incident and transmitted far-IR light, respectively. We
consider two samples; Sample-1 is a non-doped Ge wafer with a doped Ge
layer, while Sample-2 is a non-doped Ge wafer without a doped Ge
layer. Hereafter, we call the relative transmittance between the two samples
$T_{\mathrm{IR},1}/T_{\mathrm{IR},2}$ as the far-IR transmittance of
the doped Ge layer $T_{\mathrm{IR,d}}$.

We require $T_{\mathrm{IR,d}}\ge0.95$ at $T=4$ K and $\lambda=100$ $\mu$m.
To calculate $T_{\mathrm{IR,d}}$, we take into account the free-carrier
absorption and photoionization processes in the doped Ge layer with
multiple refraction at the doped layer based on \cite{hadek1985}. They
presented a theoretical analysis of the far-IR 
transmittance through ion-implanted Ge samples taking into account
free-carrier absorption in the doped Ge layer with its impurity
concentration range of $10^{16}\la n_\mathrm{d}\la10^{19}$ cm$^{-3}$.  
In their analysis, they introduced an absorption parameter $a$ in the
Fresnel coefficients to calculate the transmittance from the vacuum to
the non-doped Ge medium through the doped Ge layer and the 
reflectance from the doped Ge layer surface~\citep[see Fig.~4
of][]{hadek1985}. In our calculation, to include the photoionization
process, the absorption parameter is redefined as
$a=\mathrm{exp}(-\alpha d/2 - \mathrm{i}dn\omega/c)$, where $\alpha$,
$n$, and $c$ are a photoabsorption coefficient, the complex refractive index of Ge,
and the speed of light, respectively; \cite{hadek1985} case
corresponds to the case of $\alpha=0$. The photoabsorption coefficient
is $\alpha=\sigma n_\mathrm{d}$, where $\sigma$ is a photoionization
cross-section of a dopant. The complex refractive index depends on the infrared
conductivity $\sigma(\omega)$ = $\sigma_0/(1+\mathrm{i}\omega\tau)$,
where $\sigma_0$ and $\tau$ are the dc conductivity and a single
relaxation time, respectively~\citep{hadek1985}.

The layer thickness and doping concentration at 
$T_{\mathrm{IR,d}}$=0.95 are calculated by considering the DC
conductivity dependence on impurity concentration at 4 K in
\cite{hadek1985} and $\sigma$ assumed to be 10$^{-14}$ 
cm$^{-2}$ \citep{bratt1977, wang1986}.  
As a result, as can be seen in Fig.~\ref{fig2}, smaller thickness and
lower doping concentration than those on the solid line are required to
achieve $T_{\mathrm{IR,d}}>0.95$.      

\vspace{-5mm}  
\subsection{Resistivity}
The requirement for the resistivity is that the resistance of the doped Ge
layer ($R_\mathrm{t}$) is much lower than that of a Ge PC pixel
($R_\mathrm{d}$) for a voltage drop to become negligible at the layer.
The requirement is particularly stringent for the large-format array to
uniformly apply a detector bias voltage to pixels. Here, we consider 
an $N \times N$-pixel array that has the same electrode configuration as
that in Fig.~\ref{fig1}; the array has a common doped Ge
layer for the pixels. A common metal electrode is formed on the
doped Ge layer of pixels on the outermost circumferential
sides. At the central pixel of the array, which is farthest from the
metal electrode, the fraction of the bias voltage drop $F$ is expressed as
$F=R_\mathrm{t}/(R_\mathrm{d}+R_\mathrm{t})$. To obtain $F=0.01$, the
resistivity of the doped Ge layer $\rho_\mathrm{t}$ at 4 K should be 
$\rho_\mathrm{t} = 0.02\rho_\mathrm{d}d/ NL$, where $\rho_\mathrm{d}$
and $L$ are the resistivity of the Ge PC pixel at 4 K and the pixel
size, respectively. The experimental relation between $\rho_\mathrm{t}$
and $n_\mathrm{d}$ is obtained from \cite{fritzsche1960}. The dashed line
in Fig.~\ref{fig2} is derived from the above equation on condition that
$\rho_\mathrm{d}$, $L$, and $N$ are $6\times10^8$ $\Omega$ cm
\citep{hiromoto1989}, $5\times10^{-2}$ cm, and 128, respectively.  

\vspace{-5mm}  
\subsection{Ohmic contact}
To make an ohmic contact, a contact resistance $R_\mathrm{c}$ needs to be
sufficiently small compared with $R_\mathrm{d}$.  
\cite{sze2007} introduced the characteristic energy $E_{00}$ to
categorize a carrier transport process for metal-semiconductor
contacts. For p-type Ge with doping concentration of $\ge10^{16}$
cm$^{-3}$, $E_{00}$ is estimated to be $\ge2\times10^{-3}$ eV by applying
the hole effective mass of 0.075 $m_\mathrm{e}$ \citep{hadek1985}. 
Under the operation temperature ($T\le4$ K) of Ge PCs, $E_{00}$ is much
higher than $kT$, which means a \textit{pure} tunneling process through
a Schottky barrier dominates. In this process, higher doping
concentration is required. Also, the thickness of the doped Ge layer
should be larger than that of the depletion width $W_\mathrm{D}$ due to
the metal/Ge contact.  

As a typical pixel of Ge PCs, $R_\mathrm{d}$ is estimated to be
$2\times10^9$ $\Omega$ by using $L$ and $\rho_\mathrm{d}$ at 4 K. 
Under the tunneling process, the specific
contact resistance ($R_\mathrm{sc}$) depends strongly on
$n_\mathrm{ad}$, the carrier effective mass ($m^{*}$), and the barrier
height ($\phi_\mathrm{B}$), but virtually independent of
temperature~\citep{sze2007}. 
%\begin{equation}
%R_\mathrm{sc} \propto m^{*-1}
% \mathrm{exp}\left(\phi_\mathrm{B}\sqrt{m^* n_\mathrm{ad}^{-1}}\right)
%\label{eq1}
%\end{equation}
Based on $R_\mathrm{sc}$ for a metal/n-type Ge contact measured by
\cite{gallacher2012} ($R_\mathrm{sc}\sim2\times10^{-7}$ $\Omega$ cm$^2$ for
$n_\mathrm{ad}=3\times10^{19}$ cm$^{-3}$ and $\phi_\mathrm{B}=0.75$ eV), 
by applying effective masses of hole and
electron~\citep[0.2~$m_\mathrm{e}$,][]{kahn1955}, we roughly estimate
$R_\mathrm{sc}$ for a metal/p-type Ge contact as
\begin{equation}
R_\mathrm{sc} \sim 5\times10^{-7}
 \mathrm{exp}\left(\frac{\phi_\mathrm{B}}{0.75\ \mathrm{eV}}\sqrt{0.4
	      \left(\frac{3\times10^{19}\
	       \mathrm{cm}^{-3}}{n_\mathrm{ad}}\right)}\right)\
 \Omega~\mathrm{cm^2}.
\label{eq2}
\end{equation} 
For an Al/Ge contact, since $\phi_\mathrm{B}$ is 0.5 eV
at 4 K~\citep{thanailakis1973, sze2007}, $R_\mathrm{sc}$ is estimated to
be $\la 5\times10^3$ $\Omega$ cm$^2$ for $n_\mathrm{ad}\ge10^{16}$
cm$^{-3}$. Therefore, we obtain $R_\mathrm{c}\la 10^6$ $\Omega$ by
dividing $R_\mathrm{sc}$ by $L^2$, which is much lower than $R_\mathrm{d}$; the condition
of $R_\mathrm{c} \ll R_\mathrm{d}$ is satisfied for
$n_\mathrm{ad}\ge10^{16}$ cm$^{-3}$.  

\begin{figure}
\center{
\hspace{-20mm}
\includegraphics[scale=.60]{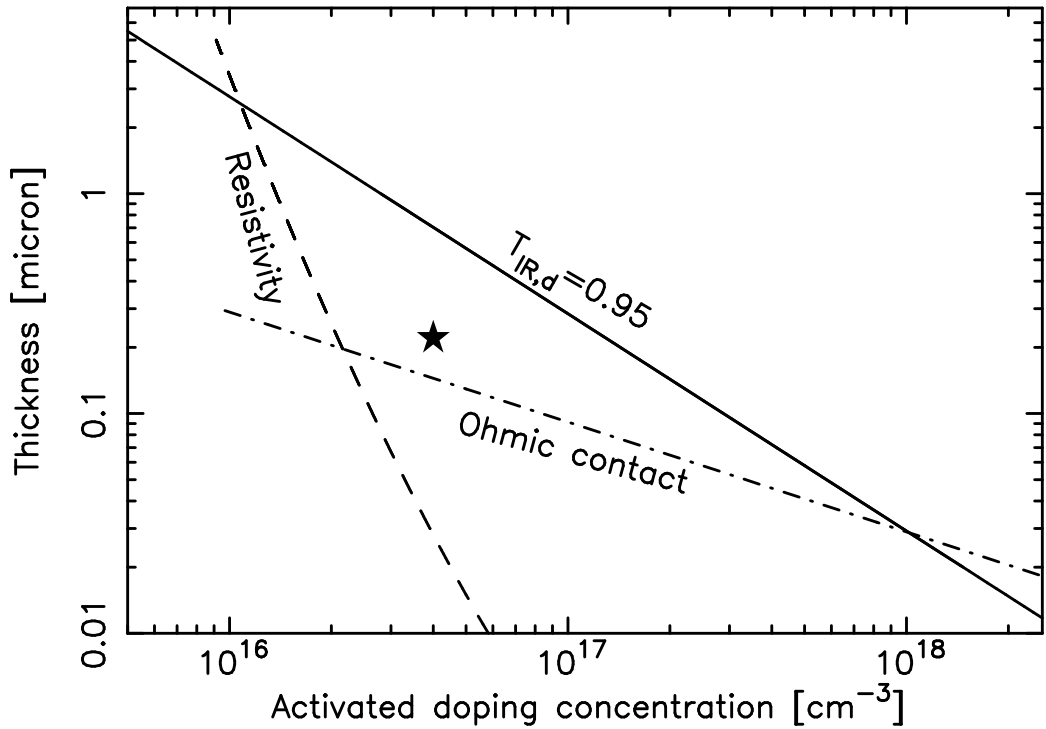}
\caption{Requirements on the thickness and doping concentration of a
 far-IR transparent electrode for Ge PCs at 4 K. The 
 solid line shows $T_\mathrm{IR,d}=0.95$ at $\lambda$ = 100
 $\mu$m. The dashed and dash-dotted lines are constraints in terms of
 resistivity and ohmic contact, respectively. The enclosed area by the
 three lines is best solutions for the parameters to fabricate on
 excellent far-IR transparent electrode. The filled star shows the
 parameters measured for our Al-doped Ge layer.} 
\label{fig2}
}
\end{figure}

In the case of $R_\mathrm{c} \ll R_\mathrm{d}$, which means a voltage drop at the
contact is negligible, $W_\mathrm{D}$ is simply expressed as
$W_\mathrm{D}=\sqrt{2 \epsilon_s V_\mathrm{bi}/qn_\mathrm{ad}}$, where
$\epsilon_s$ and $V_\mathrm{bi}$ (0.5 V at 4 K) are the permittivity of
Ge and built-in voltage, respectively. In Fig.~\ref{fig2}, the
dash-dotted line corresponds to the condition of $d=W_\mathrm{D}$ for
the Al/Ge contact.      

From the above three requirements, the enclosed area by the three lines
in Fig.~\ref{fig2} shows optimized solutions for the fabrication of the
far-IR transparent electrode for Ge PCs.

\section{Experiments}
%Requirements for far-IR transparent contact
%mott transition:($n_\mathrm{d}\sim 10^{18}$ cm$^{-3}$)

%(1) High transmittance :  lower doping concentration 
%(2) Good ohmic contact :  higher doping concentration and thicker thickness
%(3) Low resistivity    :  higher doping concentration, good clystalline

% (1) activated doping concentration : 1e16 - 1e18/cc (Mott transition)
%     <- reduce free carrier and impurity absorptions  
% (2) Small variation in concentration
%     <- variarion within ~10 %  
% (3) Thin thickness
%     <- reduce impurity absorption, ohmic contact     
% (4) Epitaxial growth
%     <- high conductance
% (5) Transition
%     <- 10 % of the thickness of the block layer ~1 um 
%     1e18 <-> 1e13 for 0.1 um
%     ion imp B (fujiwara) exp model:  0.174046904  0.0476728715  0.125656307
%     1e18 <-> 1e13 for 0.549 um
%     
%    MBE imp Al
%     exp model: 0.222825199  0.00229088683  0.0179587901
%     1e18 <-> 1e13 for 0.0264 um

\subsection{Molecular beam epitaxial growth}
Our MBE system is built inside a growth chamber, and contains
effusion cells for source materials, a substrate holder and heater, a pumping
system to achieve UHV ($10^{-10}$--$10^{-11}$ Torr), liquid N$_2$
cyropanels, and in-situ analysis tools: a reflection high energy
electron diffraction (RHEED) gun and a mass spectrometer. A RHEED
pattern on a screen and its brightness are measured by a CCD camera.

The source Ge was prepared from a non-doped Ge ingot with a
resistivity of 49--51 $\Omega$~cm, which means that its impurity
concentration is much less than the room-temperature intrinsic carrier 
concentration in Ge ($\sim10^{13}$ cm$^{-3}$). As for a dopant, since
the non-doped Ge shows a p-type conduction below $\sim100$ K by Hall
effect measurements, the dopant is selected from p-type materials for
Ge; we chose aluminum (Al, 6N purity). Crucibles for Ge and Al are
pyrolytic boron nitride (PBN). The effusion cell temperature for Ge is
set to be 1250 \degr C at which we observe the Ge beam flux of
$1.3\times10^{14}$ cm$^{-2}$ sec$^{-1}$. Since the fraction of
electrical activation of Al in Ge is $\sim10$ \% for our MBE system
($n_\mathrm{ad}\sim0.1n_\mathrm{d}$), the target Al doping 
concentration $n_\mathrm{Al}$ 
is 4--5$\times10^{17}$ cm$^{-3}$ to give wider acceptable range in
thickness (see Fig.~\ref{fig2}). The effusion cell temperature for Al is
set to be 690 \degr C, which corresponds to the Al beam flux of
$6\times 10^{9}$ cm$^{-2}$ sec$^{-1}$.
   
We prepared a mirror-polished Ge(100) substrate ($10\times9$ mm$^2$)
with thickness of 300 $\mu$m, which was obtained from the non-doped Ge
ingot. The substrate was degreased in methanol with ultrasonic for 10
min. After degreasing, 
the substrate was ultrasonically rinsed in deionized water (18
M$\Omega$~cm) for 10 min to dissolve the native Ge oxide and then
(I) dipped into HF (49 \% HF:H$_2$O =1:3) solution for 10 min. After
the process, (II) the substrate was blown dry in dry N$_2$ and then
exposed to 8 mW cm$^{-2}$ of radiation from a UV Hg arc lamp
($\lambda=185, 254$ nm) in air for 10 min to remove carbon (C) atoms and
grow a fresh Ge oxide layer. Then we repeated processes (I) and (II) but
with duration time of 70 min with UV irradiation. The cleaned 
substrate was immediately introduced through a load lock into a UHV
chamber. Prior to MBE growth, we checked carbon coverage of the
substrate surface by X-ray photoelectron spectroscopy (XPS) analysis. By
measuring intensities of Ge3d and C1s XPS lines, carbon coverage is
estimated to be 0.08 monolayers that corresponds to the value for the HF 
treatment method in \cite{sun2006}. After the XPS analysis, the substrate was
loaded into the MBE chamber. The chamber base pressure was
$\sim10^{-10}$ Torr, and increased by up to $10^{-9}$ Torr. The
substrate was heated at 600 \degr C to remove the Ge oxide layer. The
heating process was stopped when brightness contrast between a
diffraction spot and a diffuse component on the RHEED screen was
maximum. Finally, an Al-doped Ge layer was grown on the front side of
the substrate for 3 hours. During the growth, when we confirmed
the change of a RHEED pattern from 3D to 2D growth mode, the substrate
temperature was set to be 500 \degr C from 600 \degr C. The processed
sample was cut into two pieces for providing samples for optical and
electrical measurements.

\subsection{Optical and electrical measurements}
Far-IR transmittances of Sample-1 and -2 were measured at 4 K by using
Fourier transform infrared spectroscopy. Geometries of
the two samples are summarized in Table~\ref{table1}. To cool them down
to 4 K, the samples were installed into a cryostat in which there are a
wheel to rotate the samples in and out of the beam, a Si bolometer 
detector, and a preamplifier. Far-IR transmittance
measurements for both samples were performed by measuring this
transmission relative to that of a blank aperture. The reproducibility
of the measurement was 3--5 \%.   

To measure resistivity of the Al-doped Ge layer formed on the non-doped Ge 
substrate (Sample-1), four-point probe measurements were
performed with bias current of 10 $\mu$A at 4 K. Since most of carriers
in the substrate are trapped into impurity levels and hence the charged
impurity atoms are neutralized at 4 K (carrier freeze-out), the resistivity of the
substrate is expected to be much higher than that of the Al-doped Ge
layer at 4 K. Therefore, measured values can be regarded as the
resistivity of the Al-doped Ge layer itself.  

The condition that $R_\mathrm{c}$ is much lower than $R_\mathrm{d}$
should be satisfied to make an ohmic contact. If not ($R_\mathrm{c}
\approx R_\mathrm{d}$), current-voltage ($I$-$V$) curves
show Schottky diode behavior. In oder to investigate $I$-$V$
curves, we prepared Sample-3, an Al layer with thickness of 0.2 $\mu$m
was formed on the Al-doped Ge layer (see Table~\ref{table1}). The Al and
Al-doped Ge layers were then electrically split into two parts by
chemical etching. By using the two metal contact pads, we measured
$I$-$V$ curves for Sample-3 with the four-terminal method at
temperatures between 5 and 20 K.

\section{Results and Discussions}
\begin{figure}
\center{
\hspace{-20mm}
\includegraphics[scale=.60]{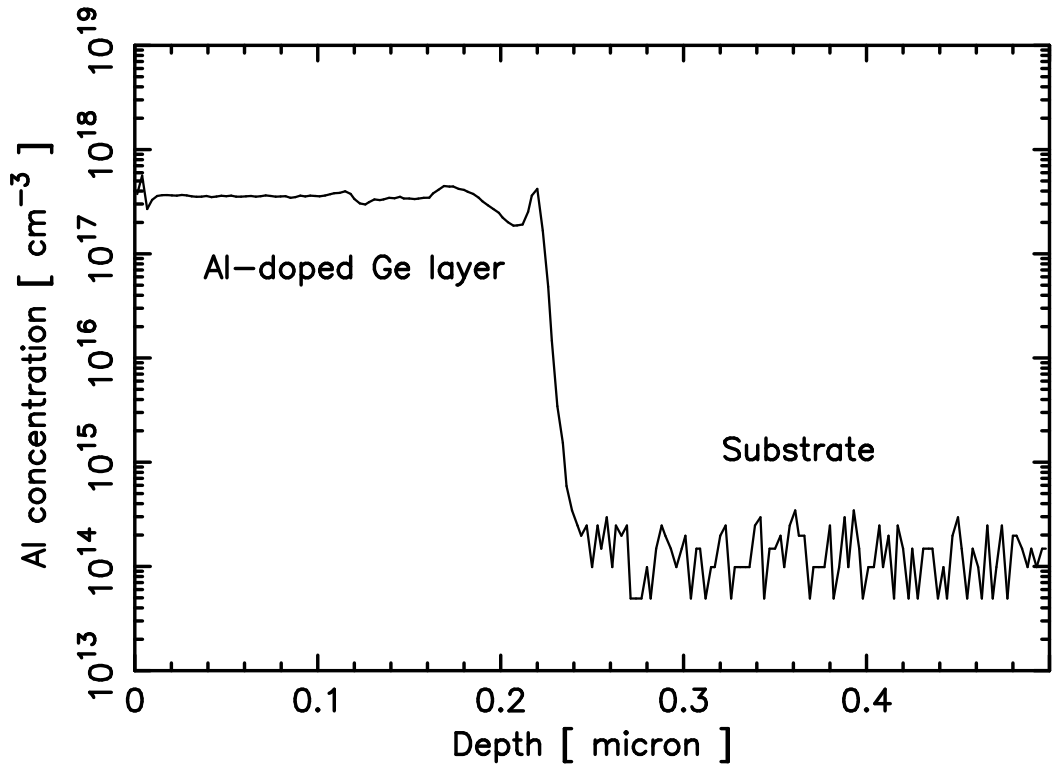}
\caption{Depth profile of the Al concentration of Sample-1. The
 detection limit of Al is $\sim10^{14}$ cm$^{-3}$.} 
\label{fig3}
}
\end{figure}

Figure~\ref{fig3} shows the depth profile of Al concentration for
Sample-1, which is obtained from a secondary ion mass spectroscopy
analysis. The detection limit of the Al concentration is about $10^{14}$ 
cm$^{-3}$. As can be seen in Fig.~\ref{fig3}, the thickness of the
Al-doped Ge layer is 0.22 $\mu$m. The Al doping concentration of
$\sim4\times 10^{17}$ cm$^{-3}$ is within the range of the target
values, whose variation is 20 \% ($3\sigma$) in the depth range of 0.02 
to 0.18 $\mu$m. Activated doping concentration is measured by a spread
resistance analysis at 300 K, and is $\sim10$ \% of the Al doping
concentration. The large variation around the depth of 0.2 $\mu$m seems
to be caused by the change in the substrate temperature from 600 \degr C
to 500 \degr C; higher substrate temperature increases the rate of
Al re-evaporation from the substrate surface and thus decreases its
concentration. The local peak of the Al concentration at the depth of
0.21--0.23 $\mu$m indicates the growth interface.

\begin{figure}
\center{
\hspace{-20mm}
\includegraphics[scale=.60]{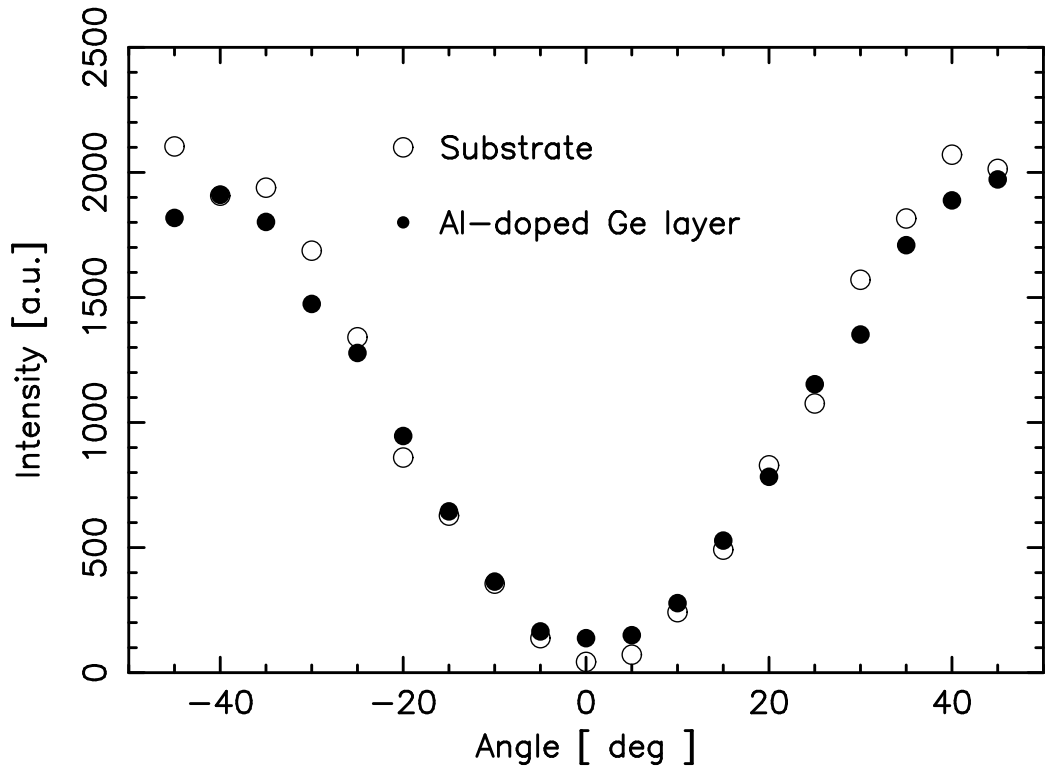}
\caption{Angle dependence of Raman line intensities for the Al-doped Ge
 layer (filled circle) and the non-doped Ge(100) substrate (open
 circle). In the $z(x,x)\bar{z}$ configuration, the polarized
 incident/scattered light forms an angle with the [100] crystal axis. }    
\label{fig4}
}
\end{figure}

At the growth interface, the profile shows an exponential
decrease with the scale length of $2\times 10^{-3}$ $\mu$m, which is
about an order of magnitude lower than that obtained from the
ion-implantation technology \citep{fujiwara2000}. In our MBE system, we
demonstrated high controllability for both doping concentration and
profile. In Fig.~\ref{fig2}, the obtained parameter set for the Al-doped
Ge layer is shown as the filled star. 
Polarized Raman spectroscopy in the $z(x,x)\bar{z}$ scattering
configuration was applied to investigate the crystal structure of the
Al-doped Ge layer. For Ge(100), if $x$, $y$, and $z$ axises are set to be 
[100], [010], and [001], respectively, the Raman line intensity should
be minimum at the angle of 0\degr between the [100] axis and the light
polarization and maximum at the angles of $\pm$45\degr.  
Figure~\ref{fig4} shows dependence of the Raman line intensity on
the angle for the Al-doped Ge layer (filled circle) and the non-doped Ge
substrate (open circle). The results confirm that the Al-doped Ge layer is 
epitaxially grown on the substrate.

Figure~\ref{fig5}(a) shows the far-IR transmittance of the
Al-doped Ge layer at 4 K, which is obtained by dividing $T_\mathrm{IR}$
of Sample-1 by that of Sample-2 as shown in
Fig.~\ref{fig5}(b). Measured data distribute around
$T_\mathrm{IR,d}=1.0$ ($40\le\lambda\le200$ $\mu$m), whereas those for ion-implanted
samples show $T_\mathrm{IR,d}=0.6$--$0.7$~\citep[e.g.][]{fujiwara2000}.
Thus, there is no significant far-IR absorption in the
Al-doped Ge layer. In Fig.~\ref{fig5}(b), $T_\mathrm{IR}$'s 
of both samples are 0.51 at $\lambda=100$ $\mu$m. By using the refractive
index of 3.7 at 4 K, which is estimated by measuring the interference
fringes around $\lambda=100$ $\mu$m, transmittance of the samples is
calculated to be 0.50, which is in good agreement with the measured
value.

The resistivity of the Al-doped Ge layer is $5\pm3$ $\Omega$ 
cm at 4 K, which is consistent with that of p-type Ge with
$n_\mathrm{ad}\sim4\times10^{16}$ cm$^{-3}$
\citep{fritzsche1960}. Although the resistivity of the
Al-doped Ge layer is much higher than that of an ion-implanted
layer \citep[$\sim10^{-2}$ $\Omega$ cm,][]{fujiwara2000}, it meets the
resistivity requirement of $\rho_t\le$38 $\Omega$ cm for $d=0.2$ $\mu$m.

At temperatures lower than $\sim$10 K, obtained $I$-$V$ curves in
Fig.~\ref{fig6}(a) show symmetry around $V=0$ mV. In particular, in the
inset that shows the $I$-$V$ curves within the range of $|V|\le400$ mV,
they show excellent symmetry and a linear dependence as indicated by the
best-fit linear regression lines.   
The $I$-$V$ curves exhibit that a linear ohmic dependence at lower voltage is
followed by nonlinear increase in slope at higher voltage. These properties are typical
$I$-$V$ curves obtained for pure Ge
crystals~\citep[$\sim10^{11}$ cm$^{-3}$,][]{teits1986}. %PCs~\citep{westervelt1985}.  
To further confirm that the $I$-$V$ curves come from the electrical
characteristic of the non-doped Ge, we investigate the relation between
the current and temperature. Under low temperatures, the non-doped Ge is
known to show p-type conduction that is attributed to residual gallium
(Ga). Since Ga acceptors in Ge have the binding energy $E_\mathrm{Ga}$ of 10.8 meV, the current is 
expected to increase with temperature in proportion to
exp($-E_\mathrm{Ga}/kT$). Figure~\ref{fig7} shows the dependence of
the current on the inverse of temperature at $V$=100 mV. At 
higher temperatures, the current exponentially increases with
$T^{-1}$ as indicated by the dashed line. Therefore, the result shows
that the electrical characteristics are dominated by those of the
non-doped Ge implying that  an excellent ohmic contact ($R_\mathrm{c}
\ll R_\mathrm{d}$) is realized below $T\sim10$ K. 

However, $I$-$V$ curves within the temperature range of 10--20 K clearly
show asymmetry around $V=0$ mV as seen in Fig.~\ref{fig6}(b). Because
the resistivity of the non-doped Ge exponentially decreases as
temperature increases from 4 K to 10 K, the ohmic contact condition for
an electrode seems to be broken ($R_\mathrm{c} \approx
R_\mathrm{d}$). In this condition, $I$-$V$ curves should be asymmetric.  
Here, we regard an equivalent circuit for one of the Al/Ge contacts as  
a circuit that consists of a Schottky diode and a resistance connected
in parallel; the equivalent circuit for the other Al/Ge contact is
considered to be a resistance. The theoretical $I$-$V$ curve expected
from the equivalent circuit model shows the dashed line in
Fig.~\ref{fig6}(b) and fits the measured data. The result means one of
the Al/Ge contacts is not the ohmic contact condition at $T\ga10$ K.    

From the fact that $R_\mathrm{c} \approx R_\mathrm{d}$ at $T\sim 10$ K,
we can roughly estimate the specific contact resistance. 
By using the measured resistivity ($\sim$10 $\Omega$ cm)
of the non-doped Ge at 10 K and considering the structure geometry of
Sample-3, $R_\mathrm{d}$ ($\approx R_\mathrm{c}$) is roughly estimated
to be $\sim1$ $\Omega$. Thus, the specific contact
resistance is $\approx10^{-1}$ $\Omega$ cm$^2$ by multiplying
$R_\mathrm{c}$ with the area of the metal electrode (0.16 cm$^2$).    
Under the tunneling process through a Schottky barrier ($kT \ll E_{00}$),
the specific contact resistance is less sensitive to temperature
\citep{sze2007}. Thus, 
$R_\mathrm{sc}\approx10^{-1}$ $\Omega$ cm$^2$ can also be applied at temperatures
below 10 K and is roughly in agreement with the expected value of
$R_\mathrm{sc}$($\sim5\times10^{-2}$ $\Omega$ cm$^2$) obtained from
Eq.(\ref{eq2}) with $n_\mathrm{ad}=4\times10^{16}$ cm$^{-3}$ and
$\phi_\mathrm{B}=0.5$ eV.   

%. On the other hand, $R_\mathrm{sc}$ is derived from
%Eq.(\ref{eq2}) with $n_\mathrm{ad}=4\times10^{16}$ $cm^{-3}$ and
%$\phi_\mathrm{B}$=0.49 eV.

%As a typical pixel of Ge PCs, we assume $L=5\times10^{-2}$ cm and
%$\rho_\mathrm{d}=10^8$ $\Omega$ cm at 4 K. Then, $R_\mathrm{d}$ is
%estimated to be $2\times10^9$ $\Omega$, which is much higher than
%$R_\mathrm{c}$ of $\approx$40 $\Omega$ derived from $\approx10^{-1}L^{-2}$.   
%Therefore, for far-IR transparent electrodes,
%$R_\mathrm{c} \ll R_\mathrm{d}$ is ensured not only for the non-doped Ge
%but also for a lightly doped ($\sim10^{14}$ cm$^3$) Ge used for PCs.    

%the current flow length between the
%metal electrodes ($\sim10^{-2}$ cm), the specific contact resistance is
%$10^{-2}$ $\Omega$ cm$^2$.   

%We demonstrated that the MBE technology surely provided the good far-IR
%transparent electrode which satisfied the three requirements. %For
%constructing a Ge PC array camera, if we assume the same structure as
%that shown in Fig.~\ref{fig1},   

\section{Conclusions}
As a far-IR transparent electrode for Ge PCs, the Al-doped Ge layer is
epitaxially formed on the non-doped Ge substrate by using the MBE
technology. The activated doping concentration and layer thickness of
the Al-doped Ge layer are $4\times10^{16}$ cm$^{-3}$ and 0.2 $\mu$m,
respectively, which are within the range of optimized solutions for the
three requirements: high far-IR transmittance, low resistivity, and ohmic 
contact. We have evaluated the optical and electrical properties of the
Al-doped Ge layer at 4 K. We obtained the far-IR transmittance of $>95$
\% within the wavelength range of 40--200 $\mu$m, while resistivity is
low enough (5 $\Omega$ cm) compared to the requirement value of 38
$\Omega$ cm for $d=0.2$
$\mu$m. We also confirm ohmic contact between the Al-doped
Ge layer and the Al electrode. We demonstrate that the MBE technology
is well applicable in fabricating an excellent far-IR transparent
electrode for Ge PCs.

%transparent electrode for extrinsic Germanium photoconductors (Ge PCs)
% at 4 K, which is fabricated by molecular beam epitaxy (MBE). 
%As a far-IR transparent electrode, an Aluminum (Al)-doped Ge layer is formed
%at well optimized doping concentration and layer thickness in terms of the
% three requirements: far-IR transmittance, resistivity, and ohmic
% contact. The Al-doped Ge layer has the far-IR transmittance of
% $>95$ \% within the wavelength range of 40-200 $\mu$m, while good
% resistivity and ohmic contact are ensured at 4 K.  
%We demonstrated that the MBE technology surely provided the good far-IR
%transparent electrode which satisfied the three requirements.

\acknowledgments
This work was supported by Grant-in-Aid for Young Scientists B
(No. 22740129) and Grant-in-Aid for Scientific Research A (No. 20244016)
and B (No. 23340053). 

%\appendix
%\section{Appendix material}

\clearpage

%% the file, uncomment the \plotone, \plottwo, and
%% If you need a layout that cannot be achieved with \plotone or
%% \plottwo, you can invoke the graphicx package directly with the
%% \includegraphics command or use \plotfiddle. For more information,
%% please see the tutorial on "Using Electronic Art with AASTeX" in the
%% documentation section at the AASTeX Web site,
%% http://www.journals.uchicago.edu/AAS/AASTeX.
%%
%% The examples below also include sample markup for submission of
%% supplemental electronic materials. As always, be sure to check
%% the instructions to authors for the journal you are submitting to
%% for specific submissions guidelines as they vary from

%% This example uses \plotone to include an EPS file scaled to
%% 80% of its natural size with \epsscale. Its caption
%% has been written to indicate that additional figure parts will be
%% available in the electronic journal.

\begin{table}
\begin{center}
\caption{Ge samples for optical and electrical measurements. \label{table1}}
\begin{tabular}{ccccc}
\tableline\tableline
Sample & Measurement & Ge substrate & Al-doped Ge layer\tablenotemark{c}&
 Al layer\tablenotemark{c}\\
\tableline
1 & OM\tablenotemark{a}, EM\tablenotemark{b} & $10\times5\times0.3$ mm$^3$&
 $10\times5\times0.0002$ mm$^3$& None\\
2 & OM & $10\times5\times0.3$ mm$^3$& None & None\\
3 & EM & $10\times4\times0.3$ mm$^3$ &
 $4\times4\times0.0002$ mm$^3$ & $4\times4\times0.0002$ mm$^3$\\
\tableline
\end{tabular}
%% Any table notes must follow the \end{tabular} command.
\tablenotetext{a}{Optical measurement}
\tablenotetext{b}{Electrical measurement}
\tablenotetext{c}{Sizes of the Al-doped Ge and Al layers for Sample-3
 are those of two split electrodes.}
\end{center}
\end{table}

\begin{figure}
\epsscale{.75}
\center{
\hspace{-20mm}
%\plottwo{fig4a.ps}{fig4b.ps}
\includegraphics[scale=.50]{fig5a.ps}\\ 
\hspace{-20mm}
\includegraphics[scale=.50]{fig5b.ps}
\caption{(a) Far-IR transmittance of the Al-doped Ge layer at 4 K and (b)
 Transmittances of Sample-1 and -2 at 4 K.}
\label{fig5}
}
\end{figure}

\begin{figure}
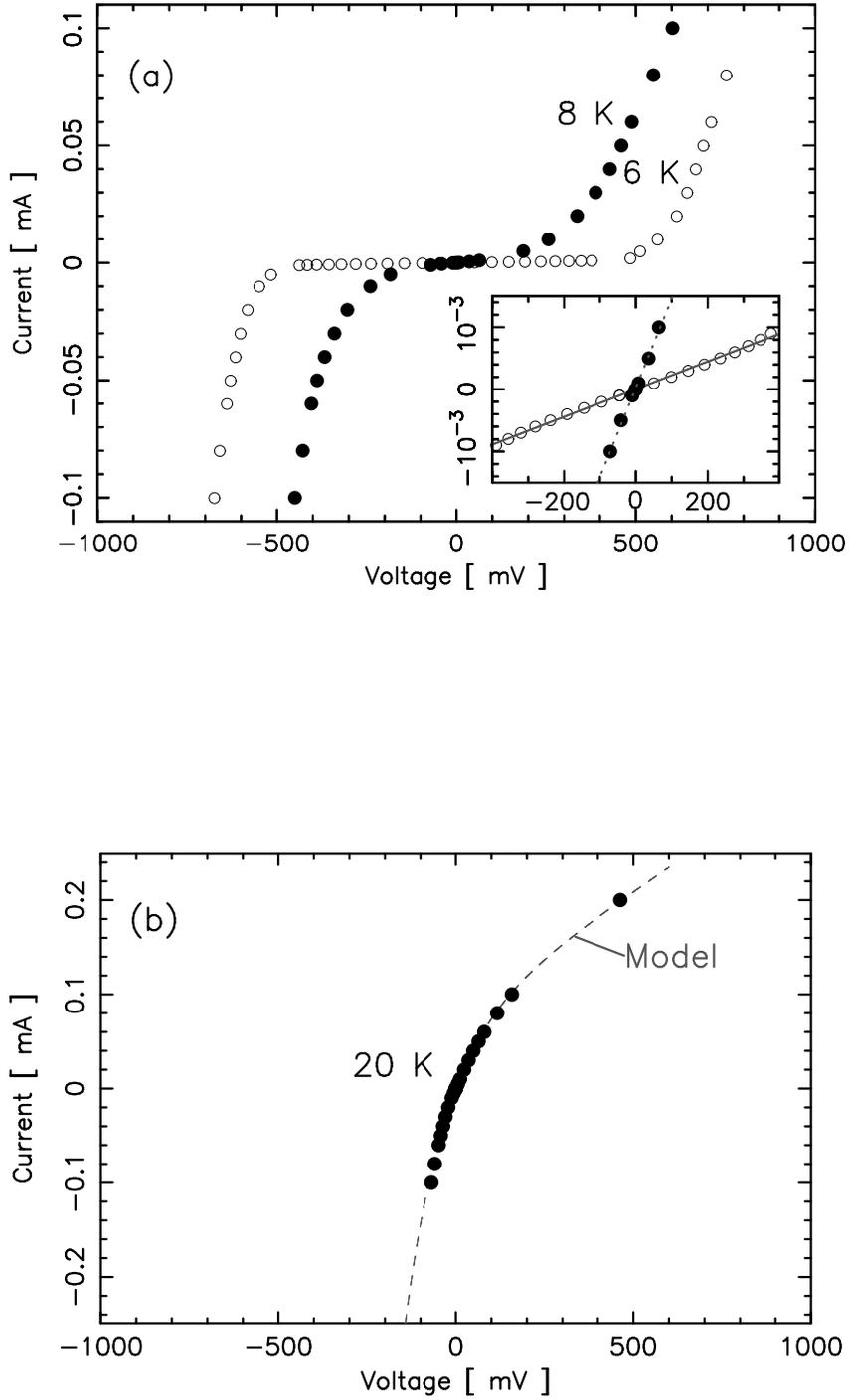

\vspace{-30mm}
\epsscale{.75}
\center{
\hspace{-20mm}
%\plottwo{fig4a.ps}{fig4b.ps}
\includegraphics[scale=.45]{fig6a.ps}\\ 
\hspace{-30mm}
\includegraphics[scale=.45]{fig6b.ps}
\caption{(a) $I$-$V$ curves of Sample-3 at 6 K (open circle) and 8 K
 (filled circle). The inset displays the enlarged view of
 the $I$-$V$ curves within the range of $|V|\le400$ mV. The solid and dotted
 lines show the best-fit linear model. (b) Same as (a) but for 20
 K. The dashed line shows the best-fit equivalent circuit model; the
 circuit for one of the Al/Ge contacts consists of a Schottky diode and
 a resistance connected in parallel, while that for the other Al/Ge contact is
 considered to be a resistance. }
\label{fig6}
}
\end{figure}

\begin{figure}
\center{
\hspace{-20mm}
\includegraphics[scale=.60]{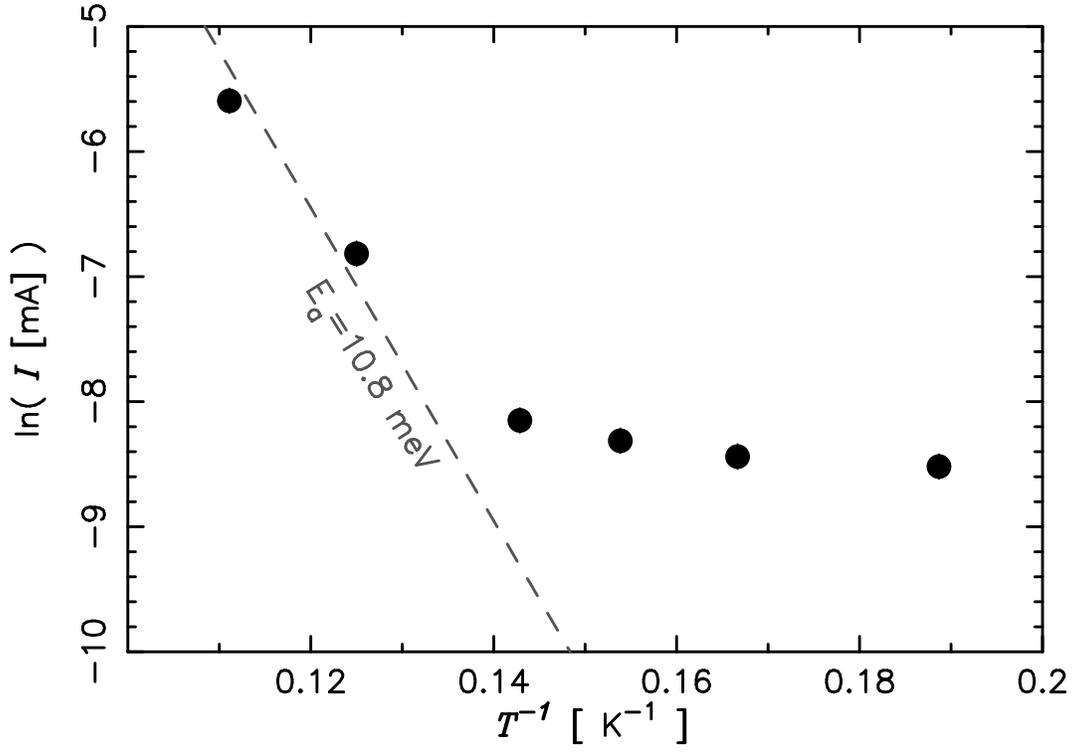}
\caption{Temperature dependence of current at $V=100$ mV for
 Sample-3. The dashed line shows $I\propto$ exp($-E_\mathrm{Ga}/kT$),
 where $E_\mathrm{Ga}$ is equal to 10.8 meV.}
\label{fig7}
}
\end{figure}

\clearpage

\clearpage

%% Tables may also be prepared as separate files. See the accompanying
%% sample file table.tex for an example of an external table file.
%% To include an external file in your main document, use the \input
%% command. Uncomment the line below to include table.tex in this
%% sample file. (Note that you will need to comment out the \documentclass,
%% \begin{document}, and \end{document} commands from table.tex if you want
%% to include it in this document.)


%% The following command ends your manuscript. LaTeX will ignore any text

\end{document}